\documentclass[12pt]{article}
\usepackage{amsmath,amssymb}
\usepackage[dvipdfmx]{graphicx}
\begin{document}

\title{Extended Landen's formulas and\\ double product theta functions}
\author{Kiyoshi \textsc{Sogo}
\thanks{EMail: sogo@icfd.co.jp}
}
\date{}

\maketitle

\begin{center}
Institute of Computational Fluid Dynamics, 1-16-5, Haramachi, Meguroku, 
Tokyo, 152-0011, Japan
\end{center}
\abstract{
For an arbitrary positive integer $p$, Landen's formula is extended to express theta function with modulus $p\tau$ by 
$p$ product of theta functions with $\tau$, which is applied to several examples. 
Next it is shown that double product of theta functions of genus $g=1$ is written by a sum of $g=2$ 
theta functions, which is a subset having a special period matrix of $\tau_{11}=\tau_{22}$. 
Several applied examples are shown, which include the cubic identity of Borwein and Borwein. 
}
%
%

\section{Introduction}
\setcounter{equation}{0}

Landen's formula which relates Jacobi's theta functions with modulus $\tau$ and $2\tau$, is given by \cite{WW}
\begin{align}
\frac{\vartheta_4(2u, 2\tau)}{\vartheta_4(u, \tau)\vartheta_3(u, \tau)}=
\text{constant}=
\prod_{n=1}^\infty\frac{(1-q^{4n})}{(1-q^{2n})^2},
\label{Landen}
\end{align}
where we set $q=e^{\pi i\tau}$. We have at least two types of proof for \eqref{Landen}, 
one is to use infinite product formulas for $\vartheta_4$ and $\vartheta_3$, 
another is to show zero points of numerator and denominator coincide, and apply Liouville's theorem.
Let us remind the former proof. By setting $z=e^{\pi i u}$, we have \cite{WW}
\begin{align}
\begin{split}
\vartheta_4(u,\tau)=\prod_{n=1}^\infty (1-q^{2n})(1-q^{2n-1}z^2)(1-q^{2n-1}z^{-2}),\\
\vartheta_3(u,\tau)=\prod_{n=1}^\infty (1-q^{2n})(1+q^{2n-1}z^2)(1+q^{2n-1}z^{-2}),
\end{split}
\end{align}
then denominator ($D$) and numerator ($N$) of the left hand side of \eqref{Landen} become
\begin{align}
D&=\prod_{n=1}^\infty(1-q^{2n})^2(1-q^{2n-1}z^2)(1-q^{2n-1}z^{-2})(1+q^{2n-1}z^2)(1+q^{2n-1}z^{-2})
\nonumber \\
&=\prod_{n=1}^\infty(1-q^{2n})^2(1-q^{4n-2}z^4)(1-q^{4n-2}z^{-4}),\\
N&=\prod_{n=1}^\infty (1-(q^2)^{2n})(1-(q^2)^{2n-1}(z^2)^2)(1-(q^2)^{2n-1}(z^2)^{-2}),
\nonumber \\
&=\prod_{n=1}^\infty (1-q^{4n})(1-q^{4n-2} z^4)(1-q^{4n-2} z^{-4}),
\end{align}
therefore we obtain $N/D$ by the right hand side of \eqref{Landen}. 

Now if we use a relation 
$\vartheta_3(u, \tau)=\vartheta_4(u+\frac{1}{2}, \tau)$, 
we can rewrite the left hand side of \eqref{Landen} by
\begin{align}
\frac{\vartheta_4(2u, 2\tau)}{\vartheta_4(u,\tau)\vartheta_4(u+\frac{1}{2}, \tau)}=\text{constant}=
\prod_{n=1}^\infty\frac{(1-q^{4n})}{(1-q^{2n})^2},
\end{align}
which makes easier to locate the zeros of denominator in the second proof. 
Since the zero point of $\vartheta_4(u, \tau)$ in unit parallelogram $(0, 1, \tau, 1+\tau)$ 
of quasi-periodicity is at $u=\frac{\tau}{2}$, zero points of denominator are at $u=\frac{\tau}{2}$ and $u=\frac{1+\tau}{2}$ which 
are also zeros of numerator.  

Then we notice that such proof can be extended further to obtain
\begin{align}
\frac{\vartheta_4(3u, 3\tau)}
{\vartheta_4(u, \tau)\vartheta_4(u+\frac{1}{3}, \tau)\vartheta_4(u+\frac{2}{3}, \tau)}=\text{constant}
=\prod_{n=1}^\infty \frac{(1-q^{6n})}{(1-q^{2n})^3},
\end{align}
which will be proved in the next section. 
The first topic of the present paper is to discuss such generalizations of Landen's formula and give some applied examples 
in \S 2. 

The second topic discussed in this paper was inspired by Kronecker's paper (1883) on his famous limit formula \cite{Kronecker}, 
which says 
two product of Jacobi's theta functions (genus $g=1$) is related with Rosenhain's \cite{Rosenhain} type theta function ($g=2$). 
We will extend Kronecker's procedure in \S 3, and find such theta functions are special {\it subset} of general 
$g=2$ theta functions, period matrix of which has the form
\begin{align}
\tau=\left(\begin{array}{cc}
\tau_0&\tau_1\\
\tau_1&\tau_0 \end{array}\right),
\end{align}
having a symmetry of $\tau_{11}=\tau_{22}\equiv \tau_0$. 
Being a subset can be understood by comparing degrees of freedom (dimension of moduli space), two versus three in full symmetry. 
Several applications are given also in \S 3. 
The last section \S 4 is devoted to summary.

. 
\section{Extensions of Landen's formula}
\setcounter{equation}{0}

\subsection{Higher order Landen's type formulas}

Let us begin this section by proving 
\begin{align}
\frac{\vartheta_4(3u, 3\tau)}
{\vartheta_4(u, \tau)\vartheta_4(u+\frac{1}{3}, \tau)\vartheta_4(u+\frac{2}{3}, \tau)}=\text{constant}
=\prod_{n=1}^\infty \frac{(1-q^{6n})}{(1-q^{2n})^3}.
\label{Landen3}
\end{align}
Since the zero points of denominator and numerator in unit parallelogram are commonly at 
$u=\frac{\tau}{2},\ \frac{\tau}{2}+\frac{1}{3},\ \frac{\tau}{2}+\frac{2}{3}$, 
the left hand side is indeed a constant. The value is computed by using
\begin{align}
\begin{split}
&\vartheta_4(u,\tau)=\prod_{n=1}^\infty (1-q^{2n})(1-q^{2n-1}z^2)(1-q^{2n-1}z^{-2}),\\
&\vartheta_4(u+\frac{1}{3},\tau)=\prod_{n=1}^\infty (1-q^{2n})(1-\omega q^{2n-1}z^2)(1-\omega^2 q^{2n-1}z^{-2}),\\
&\vartheta_4(u+\frac{2}{3},\tau)=\prod_{n=1}^\infty (1-q^{2n})(1-\omega^2 q^{2n-1}z^2)(1-\omega q^{2n-1}z^{-2}),
\end{split}
\end{align}
where we set $\omega=e^{2\pi i/3}$. Then the denominator ($D$) is given by
\begin{align}
D=\prod_{n=1}^\infty (1-q^{2n})^3\cdot \prod_{n=1}^\infty (1-q^{6n-3}z^6)(1-q^{6n-3}z^{-6}),
\end{align}
because of $(1-y)(1-\omega y)(1-\omega^2 y)=1-y^3$, and the numerator ($N$) is given by
\begin{align}
N=\prod_{n=1}^\infty (1-q^{6n})(1-q^{6n-3}z^6)(1-q^{6n-3}z^{-6}),
\end{align}
thus $N/D$ becomes the right hand side of \eqref{Landen3}. 

In the same way, generally for a positive integer $p$, we have
\begin{align}
\frac{\vartheta_4(pu,\ p\tau)}{\prod_{k=0}^{p-1}\vartheta_4(u+\frac{k}{p}, \tau)}=\text{constant}=
\prod_{n=1}^\infty\frac{(1-q^{2pn})}{(1-q^{2n})^p},
\label{Landen_p}
\end{align}
which is our goal of higher order Landen's type formula, which can be derived similarly by using
\begin{align}
\prod_{k=0}^{p-1}(1-\omega^k y)=1-y^p,\qquad(\omega=e^{2\pi i/p}).
\end{align}

By the way, the right hand side of \eqref{Landen_p} has a distinctive form, which can be expressed by 
Dedekind's eta function \cite{Apostol} defined by
\begin{align}
\eta(\tau)=q^{\frac{1}{24}}\prod_{n=1}^\infty (1-q^n). 
\label{DedekindEta}
\end{align}
It is supposed that this $q$ in \eqref{DedekindEta} is defined by $q=e^{2\pi i\tau}$ (do not be confused).  
Hence the general Landen's type formula \eqref{Landen_p} becomes
\begin{align}
\frac{\vartheta_4(pu, p\tau)}{\prod_{k=0}^{p-1}\vartheta_4(u+\frac{k}{p}, \tau)}&=
\text{constant}=
\frac{\vartheta_4(0, p\tau)}{\prod_{k=0}^{p-1}\vartheta_4(\frac{k}{p}, \tau)}
\nonumber \\
&=\prod_{n=1}^\infty \frac{(1-q^{pn})}{(1-q^{n})^p}
=\frac{\eta(p\tau)}{\eta^p(\tau)},\quad (q=e^{2\pi i\tau})
\end{align}
which may have some implications because of modular property of eta function. 

It should be noted that if we change $u\rightarrow u+\frac{1}{2}$ in \eqref{Landen_p}, we have
\begin{align}
\prod_{n=1}^\infty\frac{(1-q^{2pn})}{(1-q^{2n})^p}=
\begin{cases}
&\frac{\vartheta_3(pu,\ p\tau)}{\prod_{k=0}^{p-1}\vartheta_3(u+\frac{k}{p}, \tau)}\quad (p=\text{odd})\\
&\frac{\vartheta_4(pu,\ p\tau)}{\prod_{k=0}^{p-1}\vartheta_3(u+\frac{k}{p}, \tau)}\quad (p=\text{even}),
\end{cases}
\label{parity}
\end{align}
where the numerator of the right hand side is different depending on the parity (even-odd) of $p$. 
Further by using other transformations $u\rightarrow u+\frac{\tau}{2}$ or $u\rightarrow u+\frac{1+\tau}{2}$, 
we can change from $\vartheta_4$ to $\vartheta_1$ or $\vartheta_2$, results of which are omitted here; 
see however Example 2.2 below. 
Anyway the formula \eqref{Landen_p} using $\vartheta_4$ is the simplest.

\subsection{Applications of Landen's type formulas}
In this subsection we show some examples which are derived from Landen's type formulas.\\

\noindent
{\bf (Example 2.1)}\ Let us give a well known application of $p=2$ case, that is \eqref{Landen} by setting $u=0$,
\begin{align}
\frac{\vartheta_4(0, 2\tau)}{\vartheta_4(0, \tau)\vartheta_3(0, \tau)}=
\prod_{n=1}^\infty\frac{(1-q^{4n})}{(1-q^{2n})^2},
\label{Gauss2}
\end{align}
which gives one of iterations of arithmetic-geometric mean (AGM) found by Gauss \cite{Gauss, Borweins}, which are realized by
\begin{align}
\begin{split}
&A=\frac{a+b}{2},\quad B=\sqrt{ab},\\
&a=\vartheta_3^2(0,\tau),\quad b=\vartheta_4^2(0,\tau),\quad
A=\vartheta_3^2(0,2\tau),\quad
B=\vartheta_4^2(0,2\tau).
\end{split}
\end{align}
Here the equality \eqref{Gauss2} implies $B=\sqrt{ab}$, because
\begin{align}
&\frac{B}{\sqrt{ab}}=\frac{\vartheta_4^2(0,2\tau)}{\vartheta_4(0,\tau)\vartheta_3(0,\tau)}
=\vartheta_4(0,2\tau)\cdot\frac{\vartheta_4(0,2\tau)}{\vartheta_4(0,\tau)\vartheta_3(0,\tau)}
\nonumber \\
&=\prod_{n=1}^\infty (1-q^{4n})(1-q^{4n-2})^2\cdot \prod_{n=1}^\infty \frac{(1-q^{4n})}{(1-q^{2n})^2}
=1,
\label{AGM2}
\end{align}
since the equality 
\begin{align}
\prod_{n=1}^\infty (1-q^{4n})^2(1-q^{4n-2})^2=\prod_{n=1}^\infty (1-q^{2n})^2
\end{align}
holds.\\

\noindent
{\bf (Example 2.2)}\ Let us remind that the modular equation of degree 3 is given by (in Legendre's form) 
\cite{Jacobi, Legendre, Hardy}
\begin{align}
\left(k\ell\right)^{1/2}+\left(k'\ell'\right)^{1/2}=1
\end{align}
where we can set
\begin{align}
\begin{split}
&k=\left(\frac{\vartheta_4(0,\tau)}{\vartheta_3(0,\tau)}\right)^2,\quad 
k'=\left(\frac{\vartheta_2(0,\tau)}{\vartheta_3(0,\tau)}\right)^2,\\
&\ell=\left(\frac{\vartheta_4(0,3\tau)}{\vartheta_3(0,3\tau)}\right)^2,\quad 
\ell'=\left(\frac{\vartheta_2(0,3\tau)}{\vartheta_3(0,3\tau)}\right)^2,
\end{split}
\end{align}
therefore we have
\begin{align}
\vartheta_4(0,\tau)\vartheta_4(0,3\tau)+\vartheta_2(0,\tau)\vartheta_2(0,3\tau)=
\vartheta_3(0,\tau)\vartheta_3(0,3\tau),
\label{Modular3}
\end{align}
which is (12.8.1) of \cite{Hardy} p.218, and a proof of which will be given in Example 3.3 later.  
From \eqref{Modular3} we obtain 
\begin{align}
&\prod_{n=1}^\infty (1+q^{2n-1})^2(1+q^{6n-3})^2-\prod_{n=1}^\infty  (1-q^{2n-1})^2(1-q^{6n-3})^2\nonumber \\
&\quad =4q\prod_{n=1}^\infty (1+q^{2n})^2(1+q^{6n})^2,\qquad (q=e^{\pi i\tau}),
\label{theta_13}
\end{align}
by dropping a common factor. This identity by the way resembles 
\begin{align}
\prod_{n=1}^\infty (1+q^{2n-1})^8 - \prod_{n=1}^\infty (1-q^{2n-1})^8
=16q\prod_{n-1}^\infty (1+q^{2n})^8
\end{align}
which is also famous equality, equivalent to Jacobi's quartic theta constants identity 
$\vartheta_3^4(0,\tau)-\vartheta_4^4(0,\tau)=\vartheta_2^4(0,\tau)$. 

On the other hand, if we consider $p=3$ case of Landen's type formulas \eqref{Landen_p} and 
\eqref{parity}, we obtain 
\begin{align}
\begin{split}
\prod_{n=1}^\infty \frac{(1-q^{6n})}{(1-q^{2n})^3}
&=\frac{\vartheta_4(0,3\tau)}{\vartheta_4(0,\tau)\vartheta_4(\frac{1}{3},\tau)\vartheta_4(\frac{2}{3},\tau)}
\\
&=\frac{\vartheta_3(0,3\tau)}{\vartheta_3(0,\tau)\vartheta_3(\frac{1}{3},\tau)\vartheta_3(\frac{2}{3},\tau)}.
\\
&=\frac{4\cdot\vartheta_2(0,3\tau)}{\vartheta_2(0,\tau)\vartheta_2(\frac{1}{3},\tau)\vartheta_2(\frac{2}{3},\tau)},
\end{split}
\label{Landens3}
\end{align}
where we set $q=e^{\pi i\tau}$. 
Therefore we have for example
\begin{align}
\ell^{1/2}&=\frac{\vartheta_4(0,3\tau)}{\vartheta_3(0,3\tau)}=
\frac{\vartheta_4(0,\tau)\vartheta_4(\frac{1}{3},\tau)\vartheta_4(\frac{2}{3},\tau)}
{\vartheta_3(0,\tau)\vartheta_3(\frac{1}{3},\tau)\vartheta_3(\frac{2}{3},\tau)}
\nonumber \\
&=\frac{\genfrac{[}{]}{0pt}{}{0}{\frac{1}{2}}(\tau)\cdot\genfrac{[}{]}{0pt}{}{0}{\frac{5}{6}}(\tau)\cdot
\genfrac{[}{]}{0pt}{}{0}{\frac{7}{6}}(\tau)}
{\genfrac{[}{]}{0pt}{}{0}{0}(\tau)\cdot\genfrac{[}{]}{0pt}{}{0}{\frac{1}{3}}(\tau)\cdot
\genfrac{[}{]}{0pt}{}{0}{\frac{2}{3}}(\tau)}
=\frac{\genfrac{[}{]}{0pt}{}{0}{\frac{1}{6}}(\tau)\cdot\genfrac{[}{]}{0pt}{}{0}{\frac{3}{6}}(\tau)\cdot
\genfrac{[}{]}{0pt}{}{0}{\frac{5}{6}}(\tau)}
{\genfrac{[}{]}{0pt}{}{0}{\frac{1}{3}}(\tau)\cdot\genfrac{[}{]}{0pt}{}{0}{\frac{2}{3}}(\tau)\cdot
\genfrac{[}{]}{0pt}{}{0}{\frac{3}{3}}(\tau)}
\nonumber \\
&=\frac{\genfrac{[}{]}{0pt}{}{0}{\frac{1}{2}}}{\genfrac{[}{]}{0pt}{}{0}{1}}\cdot
\left(\frac{\genfrac{[}{]}{0pt}{}{0}{\frac{1}{6}}}{\genfrac{[}{]}{0pt}{}{0}{\frac{1}{3}}}\right)^2.
\label{modularity}
\end{align}
Here the theta constants are defined and denoted by, dropping $\tau$ in the last line,
\begin{align}
\genfrac{[}{]}{0pt}{}{\alpha}{\beta}&=\genfrac{[}{]}{0pt}{}{\alpha}{\beta}(\tau)
\equiv\vartheta\binom{\alpha}{\beta}(0, \tau)
\nonumber \\
&=\sum_{n=-\infty}^\infty{\bf e}\left[\frac{1}{2}(n+\alpha)^2\tau+(n+\alpha)(0+\beta)\right],
\label{thetaconst}
\end{align}
where we set ${\bf e}[*]=\exp[2\pi i *]$ as usual. 
To derive \eqref{modularity} we have used the property $\genfrac{[}{]}{0pt}{}{0}{\beta}=\genfrac{[}{]}{0pt}{}{0}{1-\beta}$, 
which is shown by changing $n\rightarrow -n$.

In the same way, we have for $p=5$
\begin{align}
\frac{\vartheta_4(0, 5\tau)}{\vartheta_3(0, 5\tau)}&=
\frac{\vartheta_4(0,\tau)\vartheta_4(\frac{1}{5},\tau)\vartheta_4(\frac{2}{5},\tau)
\vartheta_4(\frac{3}{5},\tau)\vartheta_4(\frac{4}{5},\tau)}
{\vartheta_3(0,\tau)\vartheta_3(\frac{1}{5},\tau)\vartheta_3(\frac{2}{5},\tau)
\vartheta_3(\frac{3}{5},\tau)\vartheta_3(\frac{4}{5},\tau)}
\nonumber \\
&=\frac{\genfrac{[}{]}{0pt}{}{0}{\frac{1}{10}}(\tau)\cdot\genfrac{[}{]}{0pt}{}{0}{\frac{3}{10}}(\tau)\cdot
\genfrac{[}{]}{0pt}{}{0}{\frac{5}{10}}(\tau)\cdot
\genfrac{[}{]}{0pt}{}{0}{\frac{7}{10}}(\tau)\cdot\genfrac{[}{]}{0pt}{}{0}{\frac{9}{10}}(\tau)}
{\genfrac{[}{]}{0pt}{}{0}{\frac{1}{5}}(\tau)\cdot\genfrac{[}{]}{0pt}{}{0}{\frac{2}{5}}(\tau)\cdot
\genfrac{[}{]}{0pt}{}{0}{\frac{3}{5}}(\tau)\cdot
\genfrac{[}{]}{0pt}{}{0}{\frac{4}{5}}(\tau)\cdot\genfrac{[}{]}{0pt}{}{0}{\frac{5}{5}}(\tau)}
\nonumber \\
&=\frac{\genfrac{[}{]}{0pt}{}{0}{\frac{1}{2}}}{\genfrac{[}{]}{0pt}{}{0}{1}}\cdot
\left(
\frac{\genfrac{[}{]}{0pt}{}{0}{\frac{1}{10}}\genfrac{[}{]}{0pt}{}{0}{\frac{3}{10}}}
{\genfrac{[}{]}{0pt}{}{0}{\frac{1}{5}}\genfrac{[}{]}{0pt}{}{0}{\frac{2}{5}}}
\right)^2,
\label{ratio_5}
\end{align}
and for $p=7$
\begin{align}
\frac{\vartheta_4(0, 7\tau)}{\vartheta_3(0, 7\tau)}&=
\frac{\genfrac{[}{]}{0pt}{}{0}{\frac{1}{2}}}{\genfrac{[}{]}{0pt}{}{0}{1}}\cdot
\left(
\frac{\genfrac{[}{]}{0pt}{}{0}{\frac{1}{14}}\genfrac{[}{]}{0pt}{}{0}{\frac{3}{14}}\genfrac{[}{]}{0pt}{}{0}{\frac{5}{14}}}
{\genfrac{[}{]}{0pt}{}{0}{\frac{1}{7}}\genfrac{[}{]}{0pt}{}{0}{\frac{2}{7}}\genfrac{[}{]}{0pt}{}{0}{\frac{3}{7}}}
\right)^2,
\label{ratio_7}
\end{align}
which continues further for odd $p$, the rule of which might be trivial.\\

\noindent
{\bf (Example 2.3)}\ In the book by Farkas and Kra \cite{FK} p.242, there is an impressive identity, 
which resembles \eqref{ratio_5}, and is apparently connected with our extended Landen's type formula \eqref{Landen_p} with $p=5$,
\begin{align}
\frac{\genfrac{[}{]}{0pt}{}{\frac{1}{5}}{1}(5\tau)}{\genfrac{[}{]}{0pt}{}{\frac{3}{5}}{1}(5\tau)}
=e^{4\pi i/5}\cdot \frac{
\genfrac{[}{]}{0pt}{}{\frac{1}{5}}{\frac{1}{5}}(\tau)\cdot
\genfrac{[}{]}{0pt}{}{\frac{1}{5}}{\frac{3}{5}}(\tau)\cdot
\genfrac{[}{]}{0pt}{}{\frac{1}{5}}{1}(\tau)\cdot
\genfrac{[}{]}{0pt}{}{\frac{1}{5}}{\frac{7}{5}}(\tau)\cdot
\genfrac{[}{]}{0pt}{}{\frac{1}{5}}{\frac{9}{5}}(\tau)}
{\genfrac{[}{]}{0pt}{}{\frac{3}{5}}{\frac{1}{5}}(\tau)\cdot
\genfrac{[}{]}{0pt}{}{\frac{3}{5}}{\frac{3}{5}}(\tau)\cdot
\genfrac{[}{]}{0pt}{}{\frac{3}{5}}{1}(\tau)\cdot
\genfrac{[}{]}{0pt}{}{\frac{3}{5}}{\frac{7}{5}}(\tau)\cdot
\genfrac{[}{]}{0pt}{}{\frac{3}{5}}{\frac{9}{5}}(\tau)},
\label{FK242}
\end{align}
where we used abbreviation \eqref{thetaconst}. 
In fact we can derive \eqref{FK242} by using \eqref{Landen_p} with $p=5$, 
and the theta function property in \cite{Shimura} p.175,
\begin{align}
&\vartheta\binom{r}{s}(u+a\tau+b, \tau)
\nonumber \\
&\qquad={\bf e}\left[-\frac{1}{2}a^2 \tau-a(u+s+b)\right]\cdot\vartheta\binom{r+a}{s+b}(u, \tau),
\end{align}
where $r=0,\ s=\frac{1}{2}$ for $\vartheta_4$, and apply to the cases $a=\frac{1}{5},\ \frac{3}{5},\  b=\frac{1}{2}$. 

Similarly we can derive next higher identities of $p=7$, such as
\begin{align}
&\frac{\genfrac{[}{]}{0pt}{}{\frac{1}{7}}{1}(7\tau)}{\genfrac{[}{]}{0pt}{}{\frac{3}{7}}{1}(7\tau)}=e^{4\pi i/7}\times
\nonumber \\
&\times\frac{
\genfrac{[}{]}{0pt}{}{\frac{1}{7}}{\frac{1}{7}}(\tau)\cdot
\genfrac{[}{]}{0pt}{}{\frac{1}{7}}{\frac{3}{7}}(\tau)\cdot
\genfrac{[}{]}{0pt}{}{\frac{1}{7}}{\frac{5}{7}}(\tau)\cdot
\genfrac{[}{]}{0pt}{}{\frac{1}{7}}{1}(\tau)\cdot
\genfrac{[}{]}{0pt}{}{\frac{1}{7}}{\frac{9}{7}}(\tau)\cdot
\genfrac{[}{]}{0pt}{}{\frac{1}{7}}{\frac{11}{7}}(\tau)\cdot
\genfrac{[}{]}{0pt}{}{\frac{1}{7}}{\frac{13}{7}}(\tau)}
{\genfrac{[}{]}{0pt}{}{\frac{3}{7}}{\frac{1}{7}}(\tau)\cdot
\genfrac{[}{]}{0pt}{}{\frac{3}{7}}{\frac{3}{7}}(\tau)\cdot
\genfrac{[}{]}{0pt}{}{\frac{3}{7}}{\frac{5}{7}}(\tau)\cdot
\genfrac{[}{]}{0pt}{}{\frac{3}{7}}{1}(\tau)\cdot
\genfrac{[}{]}{0pt}{}{\frac{3}{7}}{\frac{9}{7}}(\tau)\cdot
\genfrac{[}{]}{0pt}{}{\frac{3}{7}}{\frac{11}{7}}(\tau)\cdot
\genfrac{[}{]}{0pt}{}{\frac{3}{7}}{\frac{13}{7}}(\tau)},
\end{align}
and
\begin{align}
&\frac{\genfrac{[}{]}{0pt}{}{\frac{3}{7}}{1}(7\tau)}{\genfrac{[}{]}{0pt}{}{\frac{5}{7}}{1}(7\tau)}=e^{4\pi i/7}\times
\nonumber \\
&\times\frac{
\genfrac{[}{]}{0pt}{}{\frac{3}{7}}{\frac{1}{7}}(\tau)\cdot
\genfrac{[}{]}{0pt}{}{\frac{3}{7}}{\frac{3}{7}}(\tau)\cdot
\genfrac{[}{]}{0pt}{}{\frac{3}{7}}{\frac{5}{7}}(\tau)\cdot
\genfrac{[}{]}{0pt}{}{\frac{3}{7}}{1}(\tau)\cdot
\genfrac{[}{]}{0pt}{}{\frac{3}{7}}{\frac{9}{7}}(\tau)\cdot
\genfrac{[}{]}{0pt}{}{\frac{3}{7}}{\frac{11}{7}}(\tau)\cdot
\genfrac{[}{]}{0pt}{}{\frac{3}{7}}{\frac{13}{7}}(\tau)}
{\genfrac{[}{]}{0pt}{}{\frac{5}{7}}{\frac{1}{7}}(\tau)\cdot
\genfrac{[}{]}{0pt}{}{\frac{5}{7}}{\frac{3}{7}}(\tau)\cdot
\genfrac{[}{]}{0pt}{}{\frac{5}{7}}{\frac{5}{7}}(\tau)\cdot
\genfrac{[}{]}{0pt}{}{\frac{5}{7}}{1}(\tau)\cdot
\genfrac{[}{]}{0pt}{}{\frac{5}{7}}{\frac{9}{7}}(\tau)\cdot
\genfrac{[}{]}{0pt}{}{\frac{5}{7}}{\frac{11}{7}}(\tau)\cdot
\genfrac{[}{]}{0pt}{}{\frac{5}{7}}{\frac{13}{7}}(\tau)}.
\end{align}
Such sequences will continue for larger odd $p$, the rule of which might be obvious.


\section{Double product of theta functions}
\setcounter{equation}{0}

\subsection{Extension of Kronecker's procedure}

In this section we extend Kronecker's idea \cite{Kronecker} further, to consider double product of theta functions. 
Although Kronecker considered double product of $\vartheta_1$ functions, 
here we begin by considering double product of $\vartheta_3$ functions
\begin{align}
&\vartheta_3(x, w_1)\vartheta_3(y, w_2) \nonumber \\
&\quad =
\sum_{m,n \in {\bf Z}} {\bf e}\left[\frac{1}{2}\left(w_1m^2+w_2n^2\right)+(mx+ny)\right],
\end{align}
where inequalities $\text{Im} (w_1),\ \text{Im}(w_2)>0$ are assumed as usual. 
Let us consider a transformation
\begin{align}
\begin{split}
&m+n=M,\quad m-n=N\\
&\quad\Longleftrightarrow\quad
m=\frac{M+N}{2},\quad n=\frac{M-N}{2},
\end{split}
\label{transformation}
\end{align}
which imply integers $M, N$ are even or odd simultaneously. Then by rewriting
\begin{align}
&\frac{1}{2}\left[ w_1\left(\frac{M+N}{2}\right)^2+w_2\left(\frac{M-N}{2}\right)^2 \right]+
\left(\frac{M+N}{2}x+\frac{M-N}{2}y\right)
\nonumber \\
&=\frac{1}{2}\left[\frac{w_1+w_2}{4}(M^2+N^2)+\frac{w_1-w_2}{4}\ 2MN\right]+
\left(M\frac{x+y}{2}+N\frac{x-y}{2}\right),
\nonumber
\end{align}
we have, by writing $(M, N)=(2m, 2n)$ and $(M, N)=(2m+1, 2n+1)$, 
\begin{align}
&\vartheta_3(x, w_1)\vartheta_3(y, w_2) \nonumber \\
&=\sum_{m,n \in {\bf Z}} {\bf e}\left[\frac{1}{2}\left\{ (w_1+w_2)(m^2+n^2)+(w_1-w_2)\ 2mn \right\}\right]\times
\nonumber \\
&\times{\bf e}\left[ m(x+y)+n(x-y) \right] \nonumber \\
&+\sum_{m-\frac{1}{2},n-\frac{1}{2} \in {\bf Z}} {\bf e}\left[\frac{1}{2}\left\{ (w_1+w_2)(m^2+n^2)+(w_1-w_2)\ 2mn \right\}\right]\times \nonumber \\
&\times{\bf e}\left[ m(x+y)+n(x-y) \right] \nonumber \\
&=\vartheta\binom{00}{00}[(x+y,x-y),\tau]+\vartheta\binom{\frac{1}{2}\frac{1}{2}}{00}[(x+y,x-y),\tau], 
\label{two33}\\
&\text{with}\quad \tau=\left(\begin{array}{cc}
w_1+w_2&w_1-w_2\\
w_1-w_2&w_1+w_2\end{array}\right)\equiv\left(\begin{array}{cc}
\tau_0&\tau_1\\
\tau_1&\tau_0\end{array}\right).
\label{TauMatrix}
\end{align}
Here general two variables theta function ($g=2$) is defined by \cite{Igusa}
\begin{align}
&\vartheta\binom{\alpha_1\alpha_2}{\beta_1\beta_2}\left[ \zeta, \tau\right] 
=\sum_{m-\alpha\in{\bf Z}^2}{\bf e}\left[\frac{1}{2}\ m\tau\cdot m+
m\cdot(\zeta+\beta) \right],\\
&\text{with}\ m=(m_1, m_2),\ \alpha=(\alpha_1,\alpha_2),\ \zeta=(\zeta_1, \zeta_2),\ \beta=(\beta_1, \beta_2),\\
&\text{and}\ \tau=\left(\begin{array}{cc}
\tau_{11}&\tau_{12}\\
\tau_{21}&\tau_{22}\end{array}\right),\ \tau_{21}=\tau_{12},
\end{align}
where abbreviation of the inner product $(a_1, a_2)\cdot(b_1, b_2)=a_1b_1+a_2b_2$ is used. 
Parameters $\binom{\alpha}{\beta}$ are called {\it theta characteristics}.

By substituting $x\rightarrow x+\frac{1}{2},\ y\rightarrow y+\frac{1}{2}$ in \eqref{two33}, since  
$x+y\rightarrow x+y+1,\ x-y\rightarrow x-y$, we have 
\begin{align}
&\vartheta_4(x, w_1)\vartheta_4(y, w_2)\nonumber \\
&=\vartheta\binom{00}{00}[(x+y,x-y),\tau]-\vartheta\binom{\frac{1}{2}\frac{1}{2}}{00}[(x+y,x-y),\tau], 
\label{two44}
\end{align}
because $\vartheta_3(x+\frac{1}{2}, w)=\vartheta_4(x, w)$, and
\begin{align}
\begin{split}
&\vartheta\binom{00}{00}[(u+1, v), \tau]=+\vartheta\binom{00}{00}[(u, v),\tau],\\
&\vartheta\binom{\frac{1}{2}\frac{1}{2}}{00}[(u+1,v),\tau]=-
\vartheta\binom{\frac{1}{2}\frac{1}{2}}{00}[(u, v),\tau].
\end{split}
\end{align}
Therefore we obtain inversely
\begin{align}
&\vartheta\binom{00}{00}[(x+y, x-y),\tau]
\nonumber \\
&\quad =\frac{1}{2}\left\{\vartheta_3(x,w_1)\vartheta_3(y,w_2)+\vartheta_4(x,w_1)\vartheta_4(y,w_2)\right\},
\label{theta_00}\\
&\vartheta\binom{\frac{1}{2}\frac{1}{2}}{00}[(x+y,x-y),\tau]
\nonumber \\
&\quad =\frac{1}{2}\left\{\vartheta_3(x,w_1)\vartheta_3(y,w_2)-\vartheta_4(x,w_1)\vartheta_4(y,w_2)\right\}
\label{theta_11},
\end{align}
under the restriction that  $\tau$ has a symmetry of $\tau_{11}=\tau_{22}$ (a subset of general $\tau$'s),
\begin{align}
\tau=\left(\begin{array}{cc}
w_1+w_2&w_1-w_2\\
w_1-w_2&w_1+w_2\end{array}\right)\equiv\left(\begin{array}{cc}
\tau_0&\tau_1\\
\tau_1&\tau_0\end{array}\right),
\end{align}
which implies also
\begin{align}
w_1=\frac{1}{2}\left(\tau_0+\tau_1\right),\quad
w_2=\frac{1}{2}\left(\tau_0-\tau_1\right).
\end{align}

Now again similarly 
we have
\begin{align}
&\vartheta_2(x, w_1)\vartheta_2(y, w_2)
\nonumber \\
&=\vartheta\binom{0 \frac{1}{2}}{00}[(x+y, x-y), \tau]+
\vartheta\binom{\frac{1}{2} 0}{00}[(x+y, x-y), \tau].
\label{two22}
\end{align}
And also we have
\begin{align}
&\vartheta_1(x, w_1)\vartheta_1(y, w_2)
\nonumber \\
&=\vartheta\binom{0 \frac{1}{2}}{00}[(x+y, x-y), \tau]-
\vartheta\binom{\frac{1}{2} 0}{00}[(x+y, x-y), \tau],
\label{two11}
\end{align}
which is related with the case of Kronecker's original form.


And again we obtain inversely
\begin{align}
&\vartheta\binom{0\frac{1}{2}}{00}[(x+y, x-y), \tau]
\nonumber \\
&\quad =\frac{1}{2}\left\{\vartheta_2(x, w_1)\vartheta_2(y, w_2)+\vartheta_1(x, w_1)\vartheta_1(y, w_2)\right\},\\
&\vartheta\binom{\frac{1}{2}0}{00}[(x+y, x-y), \tau]
\nonumber \\
&\quad =\frac{1}{2}\left\{\vartheta_2(x, w_1)\vartheta_2(y, w_2)-\vartheta_1(x, w_1)\vartheta_1(y, w_2)\right\}.
\end{align}


\subsection{Applications of double product theta functions}

In this subsection we show some examples which are derived from double product theta functions.\\

\noindent
{\bf (Example 3.1)}\ If we set $x=u+\frac{1}{2},\ y=u$ and write $w_1=w_2=\tau$ in \eqref{two44}, 
by using $\vartheta_4(u+\frac{1}{2}, \tau)=\vartheta_3(u, \tau)$, we have since $\tau_0=2\tau,\ \tau_1=0$,
\begin{align}
&\vartheta_4(u, \tau)\vartheta_3(u, \tau)
\nonumber \\
&\quad =\vartheta_3(2u+\frac{1}{2}, 2\tau)\vartheta_3(\frac{1}{2}, 2\tau)
-\vartheta_2(2u+\frac{1}{2}, 2\tau)\vartheta_2(\frac{1}{2}, 2\tau)
\nonumber \\
&\quad =\vartheta_4(2u, 2\tau)\vartheta_4(0, 2\tau),
\end{align}
because of $\vartheta_2(\frac{1}{2}, 2\tau)\propto \vartheta_1(0, 2\tau)=0$. This is Landen's formula
\begin{align}
\frac{\vartheta_4(2u, 2\tau)}{\vartheta_4(u, \tau)\vartheta_3(u, \tau)}=\frac{1}{\vartheta_4(0, 2\tau)}=
\frac{\vartheta_4(0, 2\tau)}{\vartheta_4(0, \tau)\vartheta_3(0, \tau)},
\end{align}
which is \eqref{AGM2} in the previous section.\\

\noindent
{\bf (Example 3.2)}\ If we set again $w_1=w_2\equiv w$, then $\tau_0=2w,\ \tau_1=0$, and we have from \eqref{two33} and \eqref{two44},
\begin{align}
&\vartheta_3(x, w)\vartheta_3(y, w)
\nonumber \\ 
&=\vartheta_3(x+y, 2w)\vartheta_3(x-y, 2w)+\vartheta_2(x+y, 2w)\vartheta_2(x-y, 2w), 
\label{33}\\
&\vartheta_4(x, w)\vartheta_4(y, w)
\nonumber \\ 
&=\vartheta_3(x+y, 2w)\vartheta_3(x-y, 2w)-\vartheta_2(x+y, 2w)\vartheta_2(x-y, 2w),
\label{44}
\end{align}
and from \eqref{two22} and \eqref{two11},
\begin{align}
&\vartheta_2(x, w)\vartheta_2(y, w)
\nonumber \\ 
&=\vartheta_3(x+y, 2w)\vartheta_2(x-y, 2w)+\vartheta_2(x+y, 2w)\vartheta_3(x-y, 2w), \\
&\vartheta_1(x, w)\vartheta_1(y, w)
\nonumber \\ 
&=\vartheta_3(x+y, 2w)\vartheta_2(x-y, 2w)-\vartheta_2(x+y, 2w)\vartheta_3(x-y, 2w).
\end{align}
These are genus $g=1$ cases of $\tau\rightarrow 2\tau$ formulas, {\it e.g.} Igusa \cite{Igusa} p.139. 

It might be amusing to comment that \eqref{33} is a special case ($\alpha=\beta=1$) of (12.9.1) in the book 
by Hardy \cite{Hardy}. It contains however a misprint $r=1$ as the lower bound of sum that should be $r=0$, 
which has survived from the first (1940) to the recent fourth printing (2002).  
On the contrary, the original source of such relations by Tannery and Molk \cite{TanneryMolk} p.166, escaped from this error.\\

\noindent
{\bf (Example 3.3)}\ If we set $w_1=3w,\ w_2=w$ in \eqref{TauMatrix}, the Riemann's period matrix becomes
\begin{align}
\tau=\left(\begin{array}{cc}
4w&2w\\
2w&4w\end{array}\right),
\label{cubictau}
\end{align}
and the corresponding theta function becomes
\begin{align}
&\vartheta\binom{00}{00}\left[(u, v),\tau\right]=\sum_{m,n\in{\bf Z}}{\bf e}
\left[\frac{1}{2}(m,n)\tau\cdot(m,n)+(mu+nv)\right]
\nonumber \\
&\quad =\sum_{m,n=-\infty}^\infty q^{4(m^2+mn+n^2)}\ e^{2\pi i(mu+nv)},\quad
(q=e^{\pi i w}).
\end{align}
Now we notice that the choice of $w_1=3w,\ w_2=w$ is the same as Example 2.2 \eqref{Modular3} by writing $\tau=w$. 
Therefore we obtain theta constants relations at $x=y=0$ from \eqref{theta_11} and \eqref{two22}
\begin{align}
\begin{split}
&\vartheta_3(0, 3w)\vartheta_3(0, w)-\vartheta_4(0, 3w)\vartheta_4(0, w)=
2\cdot \vartheta\binom{\frac{1}{2}\frac{1}{2}}{00}[(0,0),\ \tau],\\
&\vartheta_2(0, 3w)\vartheta_2(0, w)=
\vartheta\binom{\frac{1}{2}0}{00}[(0,0),\ \tau]+\vartheta\binom{0\frac{1}{2}}{00}[(0,0),\ \tau],
\end{split}
\end{align}
which are respectively, by setting $q=e^{\pi iw}$, written by
\begin{align}
\begin{split}
&2\cdot\sum_{m, n\in{\bf Z}}q^{4\left[(m+\frac{1}{2})^2+(m+\frac{1}{2})(n+\frac{1}{2})+(n+\frac{1}{2})^2\right]},\\
&2\cdot\sum_{j, k\in{\bf Z}}q^{4\left[(j+\frac{1}{2})^2+(j+\frac{1}{2})k+k^2\right]}.
\end{split}
\end{align}
We can verify that these two are equal, because  
\begin{align}
&\left(m+\frac{1}{2}\right)^2+\left(m+\frac{1}{2}\right)\left(n+\frac{1}{2}\right)+\left(n+\frac{1}{2}\right)^2
\nonumber \\
&\qquad =\left(j+\frac{1}{2}\right)^2+\left(j+\frac{1}{2}\right) k+k^2,
\end{align}
by setting
\begin{align}
\begin{split}
&m=j+k,\quad n=-j-1
\\
&\quad\Longleftrightarrow\quad 
j=-n-1,\quad k=m+n+1.
\end{split}
\end{align}
This is one-to-one correspondence between pairs of integers $(m,\ n)$ and $(j,\ k)$. 
Hence the identity \eqref{Modular3} is proved. It should be noted that the symmetry 
$\tau_{11}=\tau_{22}=2 \tau_{12}=2 \tau_{21}$ is essential for this equality to hold. 
In this way we have
\begin{align}
\genfrac{[}{]}{0pt}{}{\frac{1}{2}\frac{1}{2}}{00}=
\genfrac{[}{]}{0pt}{}{\frac{1}{2}0}{00}=
\genfrac{[}{]}{0pt}{}{0\frac{1}{2}}{00},
\end{align}
for such period matrix.\\

\noindent
{\bf (Example 3.4)}\ 
Theta functions with the same period matrix $\tau$ of \eqref{cubictau} are also related with 
the cubic AGM by Borwein brothers \cite{Borweins, BB}, who set
\begin{align}
\begin{split}
&a(q)=\sum_{m,n=-\infty}^\infty q^{m^2+mn+n^2},\\
&b(q)=\sum_{m,n=-\infty}^\infty \omega^{m-n}\ q^{m^2+mn+n^2},\quad(\omega=e^{2\pi i/3})\\
&c(q)=\sum_{m,n=-\infty}^\infty q^{(m+\frac{1}{3})^2+(m+\frac{1}{3})(n+\frac{1}{3})+(n+\frac{1}{3})^2}.
\end{split}
\end{align}
In fact, they are expressed by our theta constants such that
\begin{align}
&a(q^4)=\vartheta\binom{00}{00}\left[(0,0),\tau\right]\equiv\genfrac{[}{]}{0pt}{}{00}{00},\\
&b(q^4)=\vartheta\binom{00}{00}\left[(\frac{1}{3},\frac{2}{3}),\tau\right]=
\vartheta\binom{00}{\frac{1}{3}\frac{2}{3}}\left[(0,0),\tau\right]\equiv\genfrac{[}{]}{0pt}{}{00}{\frac{1}{3}\frac{2}{3}},\\
&c(q^4)=\vartheta\binom{\frac{1}{3}\frac{1}{3}}{00}\left[(0,0),\tau\right]\equiv\genfrac{[}{]}{0pt}{}{\frac{1}{3}\frac{1}{3}}{00}.
\end{align}

By using \eqref{theta_00} and \eqref{theta_11}, $a(q^4)$ can be rewritten as
\begin{align}
&a(q^4)=\frac{1}{2}\left\{\vartheta_3(q^3)\vartheta_3(q)+\vartheta_4(q^3)\vartheta_4(q)\right\},
\label{BBa}
\end{align}
where we used abbreviation of theta constants $\vartheta_j(q)\equiv\vartheta_j(0, w)$ with $q=e^{\pi iw}$. 
The relation \eqref{BBa} was derived differently by Borwein brothers and Garvan \cite{BBG, Garvan}. 
They also derived
\begin{align}
b(q)=\frac{3}{2} a(q^3)-\frac{1}{2} a(q),\quad
c(q)=\frac{1}{2} a(q^{1/3})-\frac{1}{2} a(q),
\label{BBc}
\end{align}
which can be expressed also by theta constants, as is shown in the next Example 3.5.

The most remarkable discovery by Borwein brothers \cite{BB} will be {\it the cubic identity}
\begin{align}
a^3=b^3+c^3
\quad\Longleftrightarrow\quad
\genfrac{[}{]}{0pt}{}{00}{00}^3=\genfrac{[}{]}{0pt}{}{00}{\frac{1}{3}\frac{2}{3}}^3+
\genfrac{[}{]}{0pt}{}{\frac{1}{3}\frac{1}{3}}{00}^3.
\end{align}
It should be stressed that the cubic identity holds for the special period matrix of the type 
\eqref{cubictau} ($\tau_{11}=\tau_{22}=2\tau_{12}=2\tau_{21}$). 
We can expect such identity is a special case of general identity (of genus $g=2$) 
among ternary products of theta constants (with arbitrary $\tau$), which is given by
\begin{align}
3^2\cdot\genfrac{[}{]}{0pt}{}{00}{00}^3=\sum_{a',a''}{\bf e}\left(-3\ (a_1',a_2')\cdot(a_1'',a_2'')\right)\cdot
\genfrac{[}{]}{0pt}{}{a_1'a_2'}{a_1''a_2''}^3,
\label{thetaR3}
\end{align}
where theta characteristics are written by
\begin{align}
a_j',\ a_j''=\frac{0}{3},\ \frac{1}{3},\ \frac{2}{3},\qquad(j=1,2).
\end{align}
The right hand side of \eqref{thetaR3} is a sum of eighty-one ($=9^2$) terms in general.
We have $g=1$ simpler analogue of theta constants identity
\begin{align}
3\cdot\genfrac{[}{]}{0pt}{}{0}{0}^3=\sum_{a',a''}{\bf e}\left(-3 a'a''\right)\cdot
\genfrac{[}{]}{0pt}{}{a'}{a''}^3,
\end{align}
whose right hand side is a sum of nine ($=3^2$) terms; compare this with \eqref{thetaR3}. 
Such theta constants identities and theta relations are proved in  
another paper by the author. \\

\noindent
{\bf (Example 3.5)}\ 
The model by Borwein brothers \cite{BB} can be extended by making the degree of freedom increase from one to two, 
such that
\begin{align}
\begin{split}
&a(q, r)=\sum_{m, n\in{\bf Z}}q^{m^2+n^2}r^{2mn},\\
&b(q, r)=\sum_{m, n\in{\bf Z}}\omega^{m-n}q^{m^2+n^2}r^{2mn},\quad (\omega=e^{2\pi i/3})\\
&c(q, r)=\sum_{m, n\in{\bf Z}}q^{(m+\frac{1}{3})^2+(n+\frac{1}{3})^2}r^{2(m+\frac{1}{3})(n+\frac{1}{3})},
\end{split}
\end{align}
where $r^2=q$ is the original case. They are theta constants which belong to two variables theta functions with 
$q=e^{\pi i\tau_0},\ r=e^{\pi i\tau_1}$ of \eqref{TauMatrix}. 
Do not be confused since this is a different parametrization from previous Example 3.3. 

Now by using the same method of \S 3.1, it is easy to derive 
\begin{align}
\begin{split}
a(q, r)&=\vartheta_3(q_1)\vartheta_3(q_2)+\vartheta_2(q_1)\vartheta_2(q_2),\\
&=\genfrac{[}{]}{0pt}{}{0}{0}(q_1)\cdot\genfrac{[}{]}{0pt}{}{0}{0}(q_2)
+\genfrac{[}{]}{0pt}{}{\frac{1}{2}}{0}(q_1)\cdot\genfrac{[}{]}{0pt}{}{\frac{1}{2}}{0}(q_2)
\end{split}
\label{SimplestCase}
\end{align}
where we set $q_1=(qr)^2,\ q_2=(q/r)^2$. 
This reduces to Lemma 2.1 (i) (a) of \cite{BBG}, because $q_1=q^3,\ q_2=q$ if we set $r^2=q$. 
By the way, (i) (b) of \cite{BBG} is our \eqref{BBa}, 
which is also extended to
\begin{align}
a(q^4, r^4)=\frac{1}{2}\left\{\vartheta_3(q_1)\vartheta_3(q_2)+\vartheta_4(q_1)\vartheta_4(q_2)\right\},
\end{align}
where the equalities
\begin{align}
\vartheta_3(s)=\vartheta_3(s^4)+\vartheta_2(s^4),\quad 
\vartheta_4(s)=\vartheta_3(s^4)-\vartheta_2(s^4),
\end{align}
are used for $s=q_1,\ q_2$.

Now let us derive here a general formula expressing $g=2$ theta constants with period matrix \eqref{TauMatrix}, 
by a sum of products of $g=1$ theta constants. By writing $q=e^{\pi i\tau_0},\ r=e^{\pi i\tau_1}$ 
and $q_1=(qr)^2,\ q_2=(q/r)^2$ as before, we can show
\begin{align}
\genfrac{[}{]}{0pt}{}{\alpha_1\alpha_2}{\beta_1\beta_2}(q, r)
&\equiv\sum_{m, n\in{\bf Z}} q^{(m+\alpha_1)^2+(n+\alpha_2)^2} r^{2(m+\alpha_1)(n+\alpha_2)}\times
\nonumber \\
&\qquad\qquad \times{\bf e}\left[(m+\alpha_1)\beta_1+(n+\alpha_2)\beta_2\right]
\nonumber \\
&=\genfrac{[}{]}{0pt}{}{\frac{\alpha_1+\alpha_2}{2}}{\beta_1+\beta_2}(q_1)\cdot
\genfrac{[}{]}{0pt}{}{\frac{\alpha_1-\alpha_2}{2}}{\beta_1-\beta_2}(q_2)
\nonumber \\
&\qquad\qquad +\genfrac{[}{]}{0pt}{}{\frac{\alpha_1+\alpha_2}{2}+\frac{1}{2}}{\beta_1+\beta_2}(q_1)\cdot
\genfrac{[}{]}{0pt}{}{\frac{\alpha_1-\alpha_2}{2}+\frac{1}{2}}{\beta_1-\beta_2}(q_2).
\label{GeneralEquality}
\end{align}
The proof goes as follows. Under the transformation
\begin{align}
\begin{split}
&(m+\alpha_1)+(n+\alpha_2)=M+(\alpha_1+\alpha_2),\\
&(m+\alpha_1)-(n+\alpha_2)=N+(\alpha_1-\alpha_2),
\end{split}
\end{align}
which is nothing but \eqref{transformation} implying integers $M, N$ are even or odd simultaneously. 
Then by using the equalities
\begin{align}
(m+\alpha_1)^2+(n+\alpha_2)^2&=
\frac{1}{2}\left((M+\alpha_1+\alpha_2)^2+(N+\alpha_1-\alpha_2)^2 \right),\\
2(m+\alpha_1)(n+\alpha_2)&=
\frac{1}{2}\left((M+\alpha_1+\alpha_2)^2-(N+\alpha_1-\alpha_2)^2 \right),\\
(m+\alpha_1)\beta_1+(n+\alpha_2)\beta_2&=
\frac{1}{2}(\beta_1+\beta_2)(M+\alpha_1+\alpha_2) \nonumber \\
&\qquad+\frac{1}{2}(\beta_1-\beta_2)(N+\alpha_1-\alpha_2),
\end{align}
we obtain \eqref{GeneralEquality} by substitutions.

Hence we obtain, by setting $q_1=(qr)^2,\ q_2=(q/r)^2$, 
\begin{align}
&a(q, r)=\genfrac{[}{]}{0pt}{}{00}{00}(q, r)=\genfrac{[}{]}{0pt}{}{0}{0}(q_1)\cdot\genfrac{[}{]}{0pt}{}{0}{0}(q_2)
+\genfrac{[}{]}{0pt}{}{\frac{1}{2}}{0}(q_1)\cdot\genfrac{[}{]}{0pt}{}{\frac{1}{2}}{0}(q_2),\\
&b(q, r)=\genfrac{[}{]}{0pt}{}{00}{\frac{1}{3}\frac{2}{3}}(q, r)
=\genfrac{[}{]}{0pt}{}{0}{0}(q_1)\cdot\genfrac{[}{]}{0pt}{}{0}{\frac{1}{3}}(q_2)
-\genfrac{[}{]}{0pt}{}{\frac{1}{2}}{0}(q_1)\cdot\genfrac{[}{]}{0pt}{}{\frac{1}{2}}{\frac{1}{3}}(q_2),\\
&c(q, r)=\genfrac{[}{]}{0pt}{}{\frac{1}{3}\frac{1}{3}}{00}(q, r)=
\genfrac{[}{]}{0pt}{}{\frac{1}{3}}{0}(q_1)\cdot\genfrac{[}{]}{0pt}{}{0}{0}(q_2)
+\genfrac{[}{]}{0pt}{}{\frac{5}{6}}{0}(q_1)\cdot\genfrac{[}{]}{0pt}{}{\frac{1}{2}}{0}(q_2),
\end{align}
where we have used
\begin{align}
\genfrac{[}{]}{0pt}{}{0}{1}=\genfrac{[}{]}{0pt}{}{0}{0},\quad
\genfrac{[}{]}{0pt}{}{0}{-\frac{1}{3}}=\genfrac{[}{]}{0pt}{}{0}{\frac{1}{3}},\quad
\genfrac{[}{]}{0pt}{}{\frac{1}{2}}{1}=-\genfrac{[}{]}{0pt}{}{\frac{1}{2}}{0},\quad
\genfrac{[}{]}{0pt}{}{\frac{1}{2}}{-\frac{1}{3}}=\genfrac{[}{]}{0pt}{}{\frac{1}{2}}{\frac{1}{3}}.
\end{align}

These $a(q, r),\ b(q, r)$ and $c(q, r)$ does not satisfy the cubic identity $a^3=b^3+c^3$ except when $r^2=q$, 
and naturally we need some additional terms, which the author could not have identified yet, 
although certainly they are included in theta constants identity \eqref{thetaR3}, 
if we set $\tau_{11}=\tau_{22}\equiv\tau_0,\ \tau_{12}=\tau_{21}\equiv\tau_1$ and $q=e^{\pi i\tau_0},\ r=e^{\pi i\tau_1}$.

\section{Summary}
\setcounter{equation}{0}
Two topics of theta functions are considered, the first is to extend Landen's formula to higher orders, 
and the second is to express double product of genus $g=1$ theta functions by a sum of $g=2$ theta functions, and 
inversely to express a subset of $g=2$ theta functions by a sum of products of $g=1$ theta functions. 

Our result of Landen's type formulas is summarized by \eqref{Landen_p}, which is applied to
three examples, some are known results and some are new ones, generalizations of results by Farkas and Kra \cite{FK}. 

Results of double products of theta functions are given by \eqref{two33}, \eqref{two44}, \eqref{two22}, \eqref{two11} 
and their inverses, which are applied to five examples. 
Especially two of them, Example 3.4 and 3.5, are related with the problem extending the cubic 
identity by Borwein brothers \cite{BB}, which is connected to theta constants identity \eqref{thetaR3} and theta relations. 
Such problem is answered in another paper by the author, which proves general theta relations,  
{\it i.e.} identities among arbitrary $n$ products of theta functions with any genus $g$.

\end{document}